\documentclass[aps,prl,twocolumn,amsmath,amssymb,superscriptaddress,floatfix,showpacs]{revtex4}

\usepackage{graphicx}

\begin{document}
\title{Doped Singlet-Pair Crystal in the Hubbard model on the checkerboard lattice}

\author {Didier Poilblanc}
\affiliation{Laboratoire de Physique Th\'eorique, C.N.R.S. \&
Universit\'e de Toulouse, F-31062 Toulouse, France }
\author {Karlo Penc}
\affiliation{Research Institute for Solid State Physics and
Optics, H-1525 Budapest, P.O.B. 49, Hungary}
\author {Nic Shannon}
\affiliation{H.H.~Wills Physics Laboratory, University of  Bristol,
  Tyndall Avenue, Bristol BS8 1TL, UK} 
\pacs{75.10.-b, 75.10.Jm, 75.40.Mg, 74.20.Mn, 71.10.Fd}
\begin{abstract}
In the limit of large nearest--neighbor and on--site 
Coulomb repulsions, the
Hubbard model on the planar pyrochlore lattice maps, near
quarter-filling, onto a doped quantum fully packed loop model. The
phase diagram exhibits at quarter filling a novel quantum state of
matter, the Resonating Singlet-Pair Crystal, an insulating phase
breaking lattice symmetry. Properties of a few doped holes or
electrons
are investigated. In contrast to the doped quantum antiferromagnet,
phase separation is restricted to very small hopping leaving an
extended``window'' for superconducting pairing. However
the later is more fragile for large hopping than in
the case of the antiferromagnet.
\end{abstract}
\date{\today}
\maketitle

Correlated fermions and bosons on frustrated lattices
exhibit fascinating properties. For example, on the triangular
lattice, hard-core bosons with {\it nearest neighbor} (NN)
repulsion can realize a supersolid phase
exhibiting both charge order %(driven by classical frustration) 
and superfluidity~\cite{supersolid}.  %(driven by quantum fluctuations). 
In the light of recent STM experiments suggesting 
the coexistence of superconductivity and charge order~\cite{oxychloride}, 
it is interesting to ask whether a fermionic analog of a supersolid 
could be realized by doping a Mott insulator whose ground 
state spontaneously breaks lattice symmetries ?

In this Letter we study correlated fermions on the 
checkerboard lattice, a two-dimensional array of
corner-sharing tetrahedra.  
This is a two-dimensional analog of the pyrochlore lattice found in 
numerous spinel and pyrochlore materials~\cite{Review}. 
For a filling of exactly half an
electron per site, extremely strong on--site and NN repulsion select 
a macroscopically degenerate manifold of low-energy configurations
fulfilling the so-called ``ice rules'' constraint, i.e. having exactly two
particles per tetrahedron.  
These can be mapped onto the configuration
space of a 6-vertex model~\cite{6VM}.  By associating each
particle with a {\it dimer} joining the centers of the two
corner-sharing tetrahedra, 
an alternative {\it fully packed loop representation} can be obtained. 
Although the constrained quantum dynamics of bosons~\cite{bosons} and spinless
fermions~\cite{Pollmann} differ, the phase diagrams of these
models contain a rich variety of crystalline phases breaking
lattice translations and/or rotations.  In addition,
fractionalization of a doped charge $e$ into two $e/2$ components
can also appear under some
conditions~\cite{fractional,Pollmann,note1}. 

Here, we consider the realistic case of {\it spinful} fermions (electrons)
whose spin degrees of freedom play a central role. We
show that the quarter--filled ground--state (GS) is a new
insulating crystalline quantum state involving resonant {\it spin
singlet electron pairs}. We then argue, on the basis of numerical
calculations, that pairing can emerge under light
doping. Such a superconductor is expected to break lattice 
symmetries in  a similar way to the supersolid of Ref.~\cite{supersolid} 
(although with no charge ordering).
One of the most exciting applications of these ideas is to assemblies
of cold atoms in optical lattices, where models of the form 
we propose can in  principle be realized
by the appropriate tuning of atomic 
interactions~\cite{Buechler}. 

\paragraph{The model:}
Our starting point is a (fermionic) Hubbard model extended with
nearest neighbor (${\langle ij \rangle}$ in the sum) repulsion $V$ on the checkerboard lattice:
\begin{equation}
\mathcal{H}=
  -t\sum_{\langle ij \rangle}
  \left(
     f_{i\sigma}^\dagger f_{j \sigma}^{\phantom{\dagger}}+h.c.  \right) + V \sum_{\langle ij \rangle} n_{i}n_{j} +U  \sum_i n_{i\uparrow}n_{i\downarrow}
\end{equation}
using standard notation. We are interested in the limit where the 
on--site repulsion 
$U$ is very large, forbidding double occupancy. Furthermore,
we consider the limit $V\gg t$, in which case, for quarter filling 
($\langle n_i\rangle=\frac{1}{2}$), the ground state is an insulator.
Expanding about this state we obtain, to leading order, the effective Hamiltonian:
\begin{equation}
\mathcal{H}_\square=-t_2
 \sum_\square
 \left(
     f_{i\uparrow}^\dagger f_{j\downarrow}^\dagger
    -f_{i\downarrow}^\dagger f_{j\uparrow}^\dagger
  \right)
  \left(
     f_{k\downarrow}^{\phantom{\dagger}}
     f_{l\uparrow}^{\phantom{\dagger}}
    -f_{k\uparrow}^{\phantom{\dagger}}
     f_{l\downarrow}^{\phantom{\dagger}}
  \right)+h.c.
  \label{eq:Hefft2}
\end{equation}
where $t_2=\frac{2t^2}{V}$ and the summation runs over the ``empty
squares'' of the checkerboard lattice, with the sites of a given 
square counted $ikjl$.  
%$i$,$k$,$j$ and $l$ as we go around it.
$\mathcal{H}_\square$ acts on two electrons forming a singlet on the
diagonal $kl$, transferring that singlet to the (empty) perpendicular 
diagonal $ij$.%~\cite{note2}.

We also introduce a diagonal term 
which counts the squares where $\mathcal{H}_\square$ can act 
\begin{equation}
\mathcal{H}_W  = W \sum_\square
 \left(\frac{1}{2}-2  {\bf S}_{i} \cdot {\bf S}_{j}
  \right) n_i n_j (1-n_l)(1-n_k)\,.
\end{equation}
For $W \equiv t_2$ the Hamiltonian $\mathcal{H}_\square +\mathcal{H}_W$ 
becomes a sum of projectors, and we recover the physics of the
Quantum Dimer Model~\cite{RK} at the Rokhsar--Kivelson point~:  
the GS can be written exactly as an equal-weight
superposition of all zero-energy configurations~\cite{RK}. This nontrivial
property also holds in the presence of static holes ($t=0$).

Finally, we also include the 
two-site spin exchange
\begin{equation}
H_J=-J \sum_{\langle ij \rangle}\left(\frac{1}{4}-  {\bf S}_{i} \cdot
{\bf S}_{j}
  \right) n_i n_j \, ,
\end{equation}
where, for $U\gg V\gg t$, the exchange constant is given by 
$J = \frac{4t^2}{U-V} + \frac{8t^3}{V^2} + \frac{16t^3}{UV} + \frac
{16t^3}{U(U-V)}$.
%~\cite{noteJ} 
For the undoped system $J$ and $t_2$
are therefore independent parameters.
Here we consider only 
antiferromagnetic (AF) coupling, $J>0$, although a ferromagnetic
coupling could also be realized for $t<0$. 

Away from quarter filling, when $N_h$ holes (or electrons) 
are introduced, simple
counting shows that exactly $2N_h$ tetrahedra (named hereafter
``half-hole tetrahedra") should contain a single, shared electron
(or 3 shared electrons for electron doping).  
All other tetrahedra contain exactly two shared electrons, as 
before; violation of this constraint 
would lead to an energy increase of order $V$. The single hopping
term $t$ becomes then effective by moving around the locations of
these ``fractional charges''~\cite{fractional}.  We have performed
extensive Lanczos Exact Diagonalization (ED) on a periodic
(45$^o$ tilted) square cluster of $N=32$ sites in the insulator
and for a small number of doped holes or electrons~\cite{note_size}.

%%%%%%%%%%%%%%%%%%%%%%%%%%%%%%%%%%%%%%%%%
%% Figure
\begin{figure}%[h]
  \centerline{\includegraphics*[angle=0,width=0.95\linewidth]{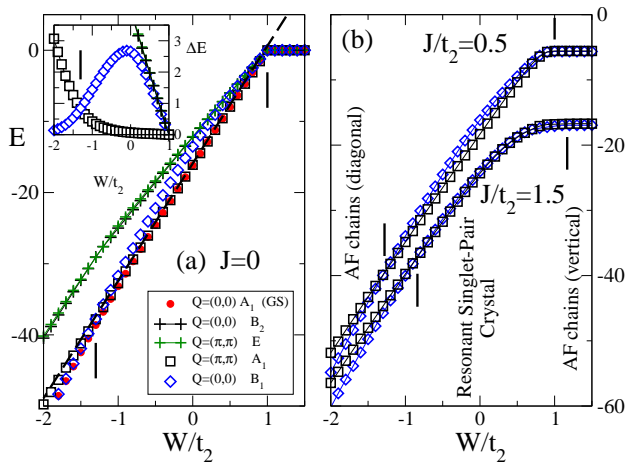}}
  \caption{\label{fig:energies}
(Color online) Energies of the 1/4-filled GS and singlet excited
states characterized by different quantum numbers (as shown) vs W
and $J=0$ (a). Inset: energy differences w.r.t. GS.
(b) Same data for excited states only and finite
values of $J$. Phase transitions (shown by vertical bars) are
signaled by the crossing between excited states (left) and by a
W-independent energy (right). The dashed line in (a) shows the
 $E^{(1)}$ perturbation result (see text).}
\end{figure}
%%%%%%%%%%%%%%%%%%%%%%%%%%%%%%%%%%%%%%%%%

\begin{figure}%[h]
  \centerline{\includegraphics*[angle=0,width=0.95\linewidth]{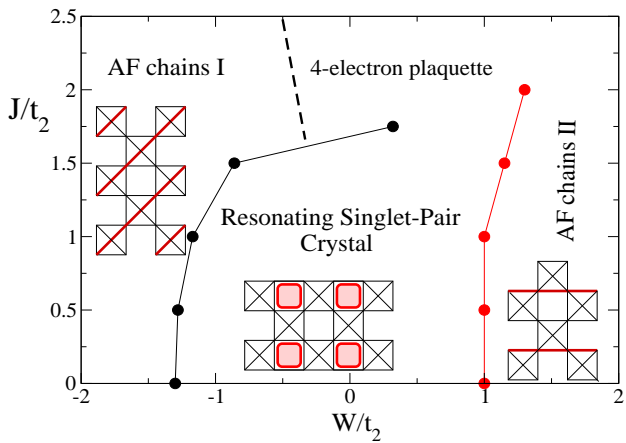}}
  \caption{\label{fig:phase_diag}
(Color online) Tentative phase diagram vs W and J. (See
description of the singlet phases in the text).}
\end{figure}

\paragraph{Phase diagram at quarter-filling:} For $J=0$, by analogy with
Ref.~\cite{bosons} and from simple analytic arguments, we
expect to have as a function of $W$: (i) for large negative $W$,
parallel spin chains; (ii) for intermediate $|W|$ values a
Resonating Singlet-Pair Crystal  (RSPC); (iii) for $W=t_2$, a 
liquid-like ``RK'' point; (iv) for $W>t_2$ a manifold of isolated states (which 
include all ferromagnetic states), all having $E=0$. This picture is
indeed supported by our ED calculations (see
Fig.~\ref{fig:energies}(a)) where the characterization of the
various phases (see Fig.~\ref{fig:phase_diag}) can be obtained
from the analysis of the low-energy spectrum. 
In the two fold
degenerate RSPC, electron pairs resonate in every second
void plaquette, breaking translational symmetry. This is
seen numerically in the collapse of a ${\bf k}=(\pi,\pi)$
${\sf A_1}$ symmetry \cite{irreps} singlet excited state onto the GS
[Fig.~\ref{fig:energies}(a)]. For sufficiently negative $W$ the
electrons order along diagonal chains to optimize the Heisenberg
exchange along the diagonals of the empty squares, hence breaking
rotation symmetry. This is seen in the ED spectrum as the
quasi--degeneracy of the (${\bf k=0}$,${\sf B_1}$) singlet and the
GS.  For $W>1$, isolated states where the plaquette resonance is 
ineffective (including the horizontal chains shown), are favored.

Note that a perturbative expansion about isolated void plaquettes
gives a very accurate estimate of the RSPC GS energy (for $J=0$). 
Already in first order --- $E^{(1)}=N(\frac{33}{64}W-\frac{1}{2}t_2)$, as shown in
Fig.~\ref{fig:energies} --- the agreement is excellent.  A comparison
to a similar expansion from disconnected diagonal chains gives an
estimate of the boundary between the two phases in full agreement
with the numerics, although it suggests that the actual critical
value shifts to more neagtive W in the thermodynamic limit.  
One should point out that (i) in contrast to the bosonic
case~\cite{bosons}, the range of stability of the plaquette phase
is much broader on the $W$ axis and (ii) as shown in
Fig.~\ref{fig:energies}(b) and Fig.~\ref{fig:phase_diag}, the RSPC
is also very stable when the exchange $J$ is included. Only above
$J/t_2\sim 1.5$, a new phase of 4-electron plaquettes oriented
along the $(1,\pm 1)$ directions is stabilized by a gain of
exchange energy (small Heisenberg loops have higher quantum
fluctuations)~\cite{note_4e}. Lastly, we stress that our RSPC at
$J/t_2<1.5$ differs from the Bond Order Wave 
%(BOW) 
realized for
$t>0$ at intermediate $V/t$ ratio (for which charge fluctuations
still occur)~\cite{Indergand} which shows two kinds of {\it
crisscrossed} plaquettes. Indeed, although both phases exhibit a
doubling of the unit cell (neither with charge ordering), they
differ in the type of resonating plaquettes. 
%We therefore expect,
%by increasing $V/t$ and $U/t$, a first order transition from the
%BOW of Ref.~\cite{Indergand} to the RSPC discussed here.

%%%%%%%%%%%%%%%%%%%%%%%%%%%%%%%%%%%%%%%%%
%% Figure
%%%%%%%%%%%%%%%%%%%%%%%%%%%%%%%%%%%%%%%%%

\paragraph{Finite doping:}
In investigating the effect of doping, we restrict to $W=0$
and consider only values of $J$ for which the RSPC is realized
at zero doping.
%Note that one expects the
%plaquette ordering to survive also at sufficiently small but {\it
%finite} doping.

\begin{figure}%[h]
  \centerline{\includegraphics*[angle=0,width=0.95\linewidth]{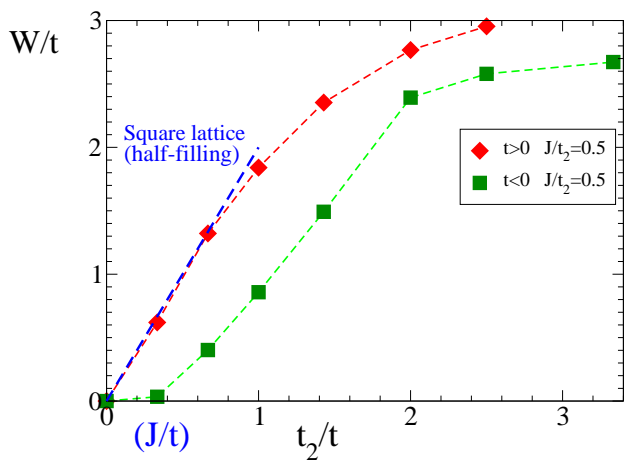}}
  \caption{\label{fig:Bandwidth_1h.eps}
(Color online) Single hole bandwidth (in units of $|t|$) as a
function of $t_2/|t|$ for a fixed (physical) ratio of $J/t_2$. For
comparison, the same quantity for the quantum AF on the square
lattice is indicated, as a function of $J/t$, by a dotted line
(blue online).}
\end{figure}

A single doped hole (or electron), as discussed above, 
is in fact split into two
mobile ``half-hole tetrahedra'' carrying one electron only (and
effective charge $e/2$)~\cite{fractional,Pollmann}. For reasons
similar to those given in Ref.~\cite{Pollmann}, one expects a confining
potential so that a small quasiparticle weight survives (in fact,
in the $t=0$ limit, the two ``half-hole tetrahedra'' even share a
common site). In order to estimate the coherent bandwidth we have
computed by ED the GS energy of a single hole for various
inequivalent ${\bf k}$ momenta~\cite{note3} and results are
reported in Fig.~\ref{fig:Bandwidth_1h.eps}. For large $t$
(compared to $t_2$ or $J$) the bandwidth is reduced by, roughly, a
factor $t_2/|t|$ (or $J/|t|$) for $t>0$, or even more for $t<0$,
and the hole becomes quite massive. 
This is very reminiscent of the behavior of a single 
hole doped  in a quantum antiferromagnet 
(shown also in Fig.~\ref{fig:Bandwidth_1h.eps} for
comparison) where the hole leaves behind a path of flipped spins
which can be healed over a short time scale $\propto 1/J$ by spin
fluctuations~\cite{Poilblanc93}. We believe a similar path along
which the plaquette order is perturbed exists also here behind the
moving hole.

 %%%%%%%%%%%%%%%%%%%%%%%%%%%%%%%%%%%%%%%%%
%% Figure
%%%%%%%%%%%%%%%%%%%%%%%%%%%%%%%%%%%%%%%%%

\begin{figure}%[h]
  \centerline{\includegraphics*[angle=0,width=0.95\linewidth]{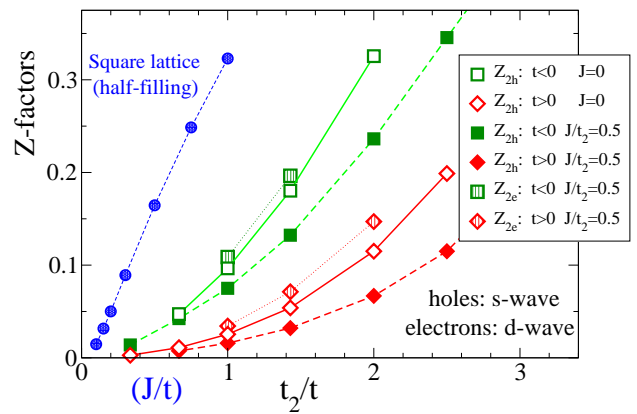}}
  \caption{\label{fig:Z2h}
(Color online) Two hole Z-factor $Z_{2h}$ versus $t_2/|t|$ for $J=0$ and
$J/t_2=0.5$. The case of two doped-electrons is also
shown ($Z_{2e}$) as hashed symbols (for $J=0.5$). For comparison 
the same quantity is shown as circles
(blue online) for two holes doped into the quantum AF on the
square lattice vs $J/t$ (data for a 26-site
cluster~\protect\cite{2holes_AF}).}
\end{figure}

As the effective Hamiltonian lowers the energy of singlet pairs,
it is natural to study the emergence of pairing interaction. 
For $t=0$ we found that two holes sit next
to each other or, more precisely, that all four ``half-hole
tetrahedra'' are fully packed around a single void plaquette in
order to minimize plaquette resonance and bond exchange energies.
The two-hole GS for $t=0$ can be very well approximated by
$|\Psi_{2h}^0\big>\simeq \Delta_s^\dagger |\Psi_0\big>$ where
$\Delta_s^\dagger$ removes two electrons along the diagonals of a
void plaquette with s-wave orbital symmetry. For
finite $t$ it is therefore convenient to define the overlap
squared $Z_{2h}=|\big<\Psi_{2h}|\Psi_{2h}^0\big>|^2$, to get an
accurate estimate of the so-called 2-hole Z-factor
$|\big<\Psi_{2h}|\Delta_s^\dagger |\Psi_0\big>|^2$. Note that
$|\Psi_{2h}^0\big>$ is now defined as a Bloch state with the same
momentum as the 2 hole GS $|\Psi_{2h}\big>$ (typically ${\bf
k}=(0,0)$) i.e. as a linear superposition of all local
(degenerate) hole pair states. Increasing $|t|$, $Z_{2h}$ should
remain finite as long as the holes stay bound. Fig.~\ref{fig:Z2h}
shows that $Z_{2h}$ is more rapidly suppressed than for two holes
in an AF~\cite{2holes_AF} (shown also in Fig.~\ref{fig:Z2h} for
comparison), suggesting that the hole pair is more fragile here
for large $|t|/t_2$ and $|t|/J$ values. A very similar behavior is
also found for two doped electrons (see Fig.~\ref{fig:Z2h}) although 
the related bound state 
has now d-wave symmetry~\cite{note4}.

 %%%%%%%%%%%%%%%%%%%%%%%%%%%%%%%%%%%%%%%%%
%% Figure
%%%%%%%%%%%%%%%%%%%%%%%%%%%%%%%%%%%%%%%%%

\begin{figure}%[h]
  \centerline{\includegraphics*[angle=0,width=0.95\linewidth]{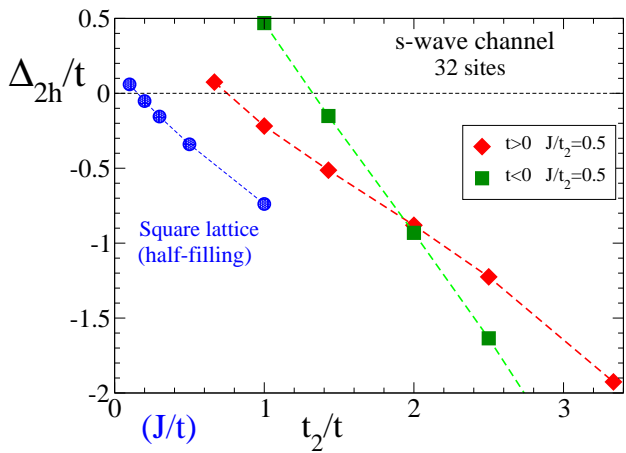}}
  \caption{\label{fig:delta_2h}
(Color online) Hole pair binding energy (in unit of $t$) vs
$t_2/|t|$ for both signs of $t$. For comparison, the case of the
doped quantum AF on the square lattice is shown (data for
$N=26$)~\protect\cite{2holes_AF}.}
\end{figure}

To get a more direct insight on the tendency of a hole pair to
break up at large $|t|/t_2$ we have computed the pair binding
energy $\Delta_{2h}=E_{2h}+E_{0h}-2E_{1h}$ shown in
Fig.~\ref{fig:delta_2h}, where $E_{mh}$ is the GS energy of the
system with $N_h=m$ holes. From its definition, negative
$\Delta_{2h}$ suggests the binding of the holes, as previously seen in 
the doped Heisenberg AF on the square lattice (in blue) --- 
even for rather large hole kinetic energy.  On the contrary, the data for the
1/4-filled checkerboard lattice do not provide evidence of binding
when $t/t_2>1$. Note however that $\Delta_{2h}$ is subject to stronger finite size
effects than $Z_{2h}$, so a weak pairing might still survive for $|t|/t_2>1$.

%%%%%%%%%%%%%%%%%%%%%%%%%%%%%%%%%%%%%%%%%
%% Figure
%%%%%%%%%%%%%%%%%%%%%%%%%%%%%%%%%%%%%%%%%

Since phase separation might compete with superconductivity, one
needs now to consider
the four-hole binding energy,
$\Delta_4=E_{4h}+E_{0h}-2E_{2h}$, from which an estimate of the
compressibility $\kappa=\frac{\partial^2 E}{\partial n_h ^2}$ at
small hole density $n_h=\frac{N_h}{N}$ can be given using
$\kappa^{-1}\simeq N\Delta_4$. Phase separation signaled by a
negative curvature of $E(n_h)$ implies, in this case, the
formation of four-hole droplets (quartets) i.e. $\Delta_4<0$. Our
numerical calculation of $\Delta_4$ on 32 sites shows that
quartets are stable when $t$ is small enough, 
the plaquette resonance and bond exchange providing 
an effective ``glue'' to bind holes together. 
Such a behavior is similar to the case of the
doped quantum AF~\cite{4holes_AF}. However, here a very small
hopping $t$ can easily suppress phase separation while affecting
pairing only slightly. Indeed, for a typical value of $J/t_2=0.5$,
$\Delta_4\simeq -0.0635\, t_2$ (meaning $\kappa <0$) for static
holes ($t=0$) while $\Delta_4\simeq -0.0042\, t_2$
($\Delta_4\simeq -0.0011\, t_2$) for $t=0.1\, t_2$ ($t=-0.1\,
t_2$) and $\Delta_4\simeq 2.471\, t_2$ ($\Delta_4\simeq 2.044\,
t_2$) for $t=0.5\, t_2$ ($t=-0.5\, t_2$). We therefore expect a
paired state when $t/t_2>0.1$.

Lastly, we would like to draw possible scenarios at finite doping:
(i) If plaquette ordering survives at sufficiently small doping
(plausible since the RSPC is ``protected'' by a finite gap), one
expects a superconducting phase which inherits broken translational
symmetry from its parent Mott insulator (the RSPC).   
This can be viewed as a new type of supersolid 
with {\it no charge modulation}~\cite{note5}; 
(ii) Alternatively, hole pairs could arrange along 
domain walls between out-of-phase
RSPC regions. This scenario is however unlikely, providing
no obvious gain of kinetic energy w.r.t the supersolid.

To summarize, on the checkerboard lattice, at quarter-filling (one
electron per two sites) and in the limit of large NN repulsion
(enforcing the ``ice rule'' constraint on every tetrahedron), the
Hubbard model exhibits a robust insulating Resonating Singlet-Pair 
Crystal with a uniform average charge. We provide arguments for
Cooper pair formation and argue in favor of a supersolid
phase in an extended
region of parameters, phase separation being confined only at very
small hole kinetic energy. However, hole-Cooper-pairs seem to be
more fragile for large hopping than in the doped antiferromagnet on
the square lattice.

%Acknowledgements
We thank P.~Fulde and F.~Pollmann for numerous discussions. D.P.
thanks the {\it Agence Nationale de la Recherche} (ANR) and 
IDRIS (Orsay, France) for support. K.P. is
grateful for the support of the Hungarian OTKA Grants
No. T049607 and No. K622800.  N.S. acknowledges support
under EPSRC Grant No. EP/C539974/1.


\begin{thebibliography}{99}
\bibitem{note1} In general, the two components are bound
by a shallow ``string'' potential.
\bibitem{irreps} A reduced Brillouin zone (BZ) scheme is used (two-site unit
  cell). The states are classified according to the 
${\sf A_1}$, ${\sf A_2}$, ${\sf B_1}$, ${\sf B_2}$, and ${\sf E}$ 
irreducible representation of the point group $D_4$ (origin at the center
of an empty square).
\bibitem{note_4e} For $W>0$ ($W<0$) an out-of-phase (in-phase) arrangements of
these diagonal lines of 4-electron plaquettes is realized to
maximize (suppress) $t_2$ resonances.
\bibitem{note4} 
For e.g. $|t|=t_2$ and $J/t_2=0.5$, the 2-doped electron (2-hole) GS
energies are $-27.428$ ($-25.935$) and $-24.704$ ($-23.220$), in units 
of $t_2$, for $t>0$ and $t<0$ respectively.
\end{thebibliography}
\end{document}